\documentclass[journal]{IEEEtran}

\usepackage[noadjust]{cite}

\usepackage{tikz,pgfplots}
\pgfplotsset{compat=1.14}
\usetikzlibrary{shadows,arrows,shapes,calc,positioning,decorations.pathreplacing,fit}
\usepgfplotslibrary{colorbrewer}
\usepackage{overpic}
\usepackage{blindtext}
\usepackage{graphicx}
\usepackage{bm}
\usepackage{color}
\usetikzlibrary{spy}
\usepackage{color}
\newcommand{\correction}[1]{{\color{black} #1}}

\usepackage{amsmath,amssymb}

\usepackage{algorithm}
\usepackage[]{algpseudocode}

\begin{document}
\bstctlcite{MyBSTcontrol}

\title{Packet Timescale Wavelength Switching Enabled by Regression Optimisation}

\author{Thomas~Gerard,~\IEEEmembership{Student Member,~IEEE,}
        Hubert~Dzieciol,~\IEEEmembership{Student Member,~IEEE,}
        Joshua~Benjamin,~\IEEEmembership{Student Member,~IEEE,}
        Kari~Clark,~\IEEEmembership{Student Member,~IEEE,}
        Hugh~Williams,
        Benn~Thomsen,~\IEEEmembership{Member,~IEEE,}
        Domani\c c~Lavery,~\IEEEmembership{Member,~IEEE,}
        and~Polina~Bayvel,~\IEEEmembership{Fellow,~IEEE}
\thanks{This work was funded by United Kingdom (UK) Engineering and Physical Sciences Research Council (EPSRC) Programme Grant TRANSNET (Transforming networks - building an intelligent optical infrastructure), EP/R035342/1. T. Gerard, H. Dzieciol and K. Clark are supported by PhD studentships from Microsoft Research. H. Dzieciol and J.  Benjamin are supported by the EPSRC. D. Lavery is supported by the Royal Academy of Engineering under the Research Fellowships scheme. }
\thanks{T. Gerard, H. Dzieciol, J. Benjamin, K. Clark, D. Lavery and P. Bayvel are with the Optical Networks Group, Department of Electronic and Electrical Engineering, University College London, London WC1E 7JE, U.K. (e-mail: \{thomas.gerard.15, hubert.dzieciol.18, joshua.benjamin.09, kari.clark.14, p.bayvel, d.lavery\}@ucl.ac.uk). }
\thanks{H. Williams and B. Thomsen are with Microsoft Research, 21 Station Road, Cambridge, CB1 2FB, U.K. (email: \{hughwi, benn.thomsen\}@microsoft.com}.
\thanks{Copyright (c) 2019 IEEE. Personal use of this material is permitted.  However, permission to use this material for any other purposes must be obtained from the IEEE by sending a request to pubs-permissions@ieee.org.}
}

\markboth{PREPRINT}%
{Shell \MakeLowercase{\textit{et al.}}: Bare Demo of IEEEtran.cls for IEEE Journals}

\maketitle

\begin{abstract}
A linear regression algorithm is applied to a digital-supermode distributed Bragg reflector laser to optimise wavelength switching times. The algorithm uses the output of a digital coherent receiver as feedback to update the pre-emphasis weights applied to the laser section currents. This permits in-situ calculation without manual weight adjustments. The application of this optimiser to a representative subsection of channels indicates this commercially available laser can rapidly reconfigure over 6.05~THz, supporting 122 channels, in less than 10~ns.
\end{abstract}

%
\IEEEpeerreviewmaketitle

\section{Introduction}

\IEEEPARstart{A}{ll} optical switch architectures can provide flat, energy efficient, low latency networking for cloud datacentres. 
Tuneable lasers are a key building block in many of these designs, applied within burst mode transmitters, tuneable filters, wavelength converters and burst mode receivers  \cite{Yin2013,Ueda2016Nagoya,Funnell2017,Sato2018,Benjamin2019}. 
\correction{Increasing the supported bandwidth of a tuneable laser can directly improve the scalability of wavelength routed networks \cite{Sato2018,Benjamin2019}, or allow for greater modulation bandwidth per channel. Furthermore, the laser must be able to switch between those channels within nanoseconds. This minimises non-transmission time \cite{Clark2018,Shi2019}, and maximises network throughput by permitting packet granularity reconfiguration \cite{Benjamin2019}. }

Integrated semiconductor lasers can be fast-switched using pre-emphasis, where the currents applied to the laser briefly overshoot their target values \cite{Rigole97}. 
The application of pre-emphasis to a custom designed, sub-C band (3.2~THz) sampled-grating distributed Bragg
reflector (DBR) laser has successfully demonstrated any-to-any tuning times of $<$5~ns \cite{Simsarian2006}. However, the extension of this technique to the commercially available digital supermode DBR (DS-DBR) laser, tested over 4.45~THz, resulted in a worst case tuning time of 88~ns \cite{Funnell2017}. This is due to the more complex multi-section laser design, the need for multi-sample pre-emphasis, and large swings in the lasing current gratings ($\sim$60~mA) \cite{Ward2005}. Furthermore, both these demonstrations required manual tuning of pre-emphasis values, limiting the practical scalability of this approach.  


In this work, we reduce the wavelength switching time of the DS-DBR laser by applying a linear regression algorithm to each channel switch combination. This algorithm was able to iteratively optimise the required pre-emphasis weights, without the need for manual calibration.   
We applied the algorithm to a selection of the most challenging channel combinations, covering the widest extremes of frequency, current and lasing sections.
Through any-to-any testing on this subset of channels, we demonstrate a 36\% increase in bandwidth and a factor of 9.1 improvement in switching time compared to \cite{Funnell2017}. 


\section{Experimental Setup}

\begin{figure}
  \centering
  \noindent\includegraphics[width=.64\linewidth]{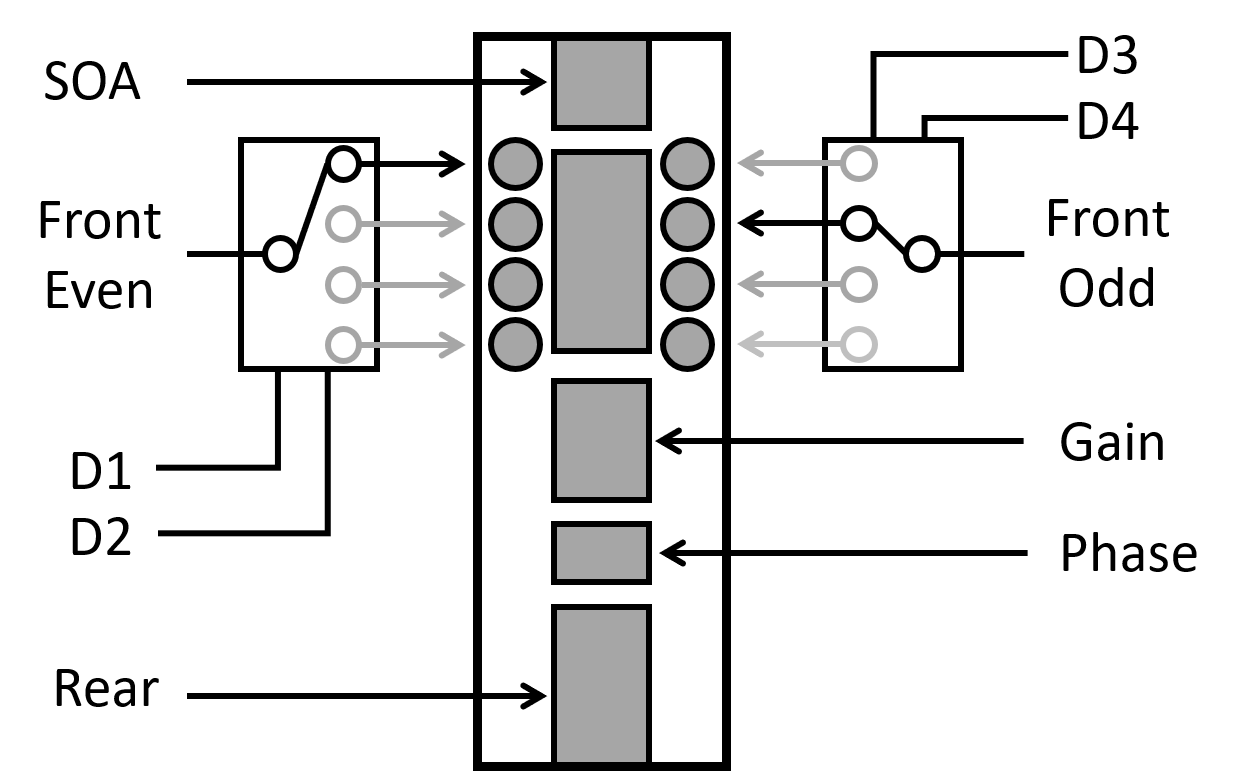}
\caption{DS-DBR laser with its 12 driving sections. Fast synchronous arbitrary current sources are needed for the Front Even, Front Odd, Rear and Phase sections. Digital signals D1-D4 are used to route the front drive currents.}
\label{fig:laser}
\vspace{-2mm}
\end{figure}

Fig.~\ref{fig:laser} shows a diagram of the Oclaro (now Lumentum) DS-DBR laser, composed of 12 driving sections. The SOA and gain section were each driven with constant current supplies of 85~mA and 140~mA respectively. 
The 8 chirped front gratings are operated in pairs. Pair $i$ is set by scaling front grating $i+1$ from 0 to 5~mA, while grating $i$ is held at 5~mA (for integer values of $i$ where $1\leq i \leq 7$); for details see \cite{Ward2005}. The laser was held at 25$^{\circ}$C using a temperature controller.

To achieve fast switching across the entire DS-DBR laser bandwidth, synchronous fast dynamic current sources are required for the rear (0-60~mA), phase (0-12~mA), and 8 front gratings (0-5~mA). In this experiment, we used 250~MS/s arbitrary waveform generators (AWG)s with 125~MHz bandwidth as fast voltage supplies. Detailed IV measurements were used to map supplied voltages to required currents. 
\correction{To reduce the required number of AWG channels, we used a digital-select analogue-route electrical multiplexer to route one AWG channel (e.g. Front Even in Fig.~\ref{fig:laser}) among the laser’s four even-numbered front gratings.  The multiplexer requires two digital selectors, D1 and D2, to select one of the four possible outputs using binary signals 00, 01, 10 or 11. A symmetric scheme is used to route Front Odd using D3 and D4. This setup allows the full laser bandwidth to be accessed using just 4 AWG channels (Front Even, Front Odd, Rear and Phase) and 4 binary channels (D1-D4). For simplicity, in our experiment all 8 of these signals were supplied by the same AWG.} 

\begin{figure}
  \centering
  \noindent\includegraphics[width=.65\linewidth]{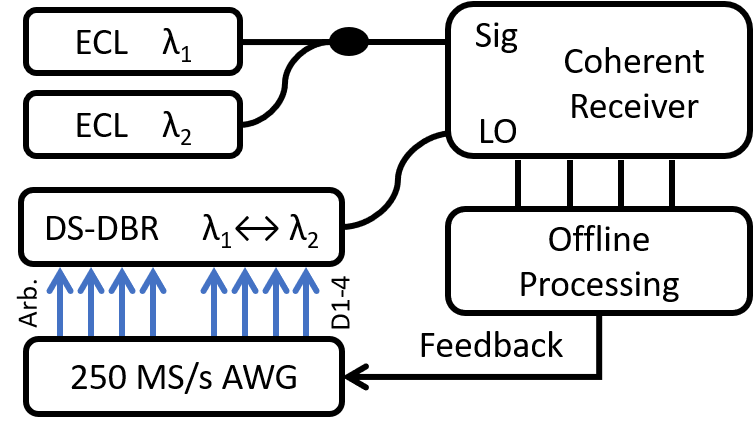}
\caption{\correction{Experimental setup. The AWG drives the DS-DBR laser with square voltage waves, inducing wavelength switching. Four 12-bit channels (Arb.) and four binary channels (D1-4) are used to access the full laser bandwidth. The external cavity lasers beat with the start and end wavelengths.}}
\label{fig:setup}
\vspace{-2mm}
\end{figure}

The setup illustrated in Fig.~\ref{fig:setup} was used to measure and improve the DS-DBR laser wavelength switching speed. To switch between two target wavelengths ($\lambda_1$ to $\lambda_2$), the DS-DBR laser was driven with 5~MHz square voltage waves from the AWGs, providing 100~ns bursts on each wavelength. Pre-emphasis was applied to the start of each burst to accelerate wavelength settling time \cite{Rigole97}. 
The laser output was passed to the local oscillator (LO) port of a coherent receiver. Two widely-tuneable (1400-1600~nm) external cavity lasers (ECL)s were set to $\lambda_1$ and $\lambda_2$, respectively, and passed to the signal port of the coherent receiver. This permitted the characterisation of the DS-DBR laser's dynamic frequency offset when switching from $\lambda_1$ to $\lambda_2$. The coherent receiver outputs were received on a 50~GS/s oscilloscope with 22~GHz bandwidth, set to capture 160,000 samples (32 bursts). These were downloaded for analysis and feedback generation. 
\correction{This measurement system could be included as a calibration node within a fast switching intra-datacentre network to share the equipment cost across many transceivers.}

\section{Regression Algorithm}

Here we propose an iterative optimisation algorithm that automatically calculates the multi-sample pre-emphasis weights needed to achieve fast switching. This optimiser uses regression to update the applied pre-emphasis sample weights based on the instantaneous frequency offset measured using the coherent receiver.  
A high level description of the pre-emphasis update procedure is shown in Algorithm~\ref{alg:filter}. The default voltage samples are represented by $\bm{x}$. A Leven frequency offset estimator \cite{Leven2007} was used to measure the instantaneous frequency offset, $\bm{e}$, as the DS-DBR laser settled at $\lambda_2$. This served as the error term for the optimiser. A vector of pre-emphasis weights, $\bm{h}$, of length $K$ was defined relative to the change in voltage needed to switch between $\lambda_1$ and $\lambda_2$ ($\Delta V$). 
To initalise the sample weights, 10 seed values for $\bm{h}$ were tested and the weights that brought $\lambda_2$ into the receiver bandwidth earliest were selected. 
Each sample weight $\bm{h}(k)$ ($k=1,2,...,K$) was updated individually using $\bm{e}(k)$, which corresponds to the error integrated over the time window, or `bin', that sample weight $k$ is applied (4~ns). If no signal is measured within a bin, the error is set to $\pm25$~GHz with the sign inferred from the following bin. Using this bin-based feedback method allowed individual weights to push past positions of local minima where changes in $\bm{h}(k)$ increased $\bm{e}(k+1)$, $\bm{e}(k+2)$, and so on. The updated voltage values, $\bm{y}$, were applied to the AWGs and the process repeated for 10 updates.

\begin{algorithm}
\caption{Pre-emphasis Optimiser}\label{euclid}
\begin{algorithmic}[1]
\State $\bm{x}=$ square waveform vector 
\For{$u=1$ to 10} 
    \State $\bm{y}=\bm{x} + \Delta V\bm{h}$
    \State \textbf{Upload} $\bm{y}$ to DS-DBR
    \State \textbf{Measure} instantaneous frequency offset
    \State $\bm{e} = $ instantaneous frequency offset
    \For{$k=1$ to $K$}
        \State $\bm{h}(k) = \bm{h}(k)-\mu \bm{e}(k)\bm{x}(k)$
    \EndFor{}
\EndFor{}
\end{algorithmic}
\label{alg:filter}
\end{algorithm}

As the laser rear grating requires the largest current swings, the optimiser was initially applied to just the rear drive voltages. For rising voltage transitions, a two sample ($K=2$) weight vector was used. For falling edges, $K=4$ was used. 

\begin{figure}[bh!]
  \centering
  \input{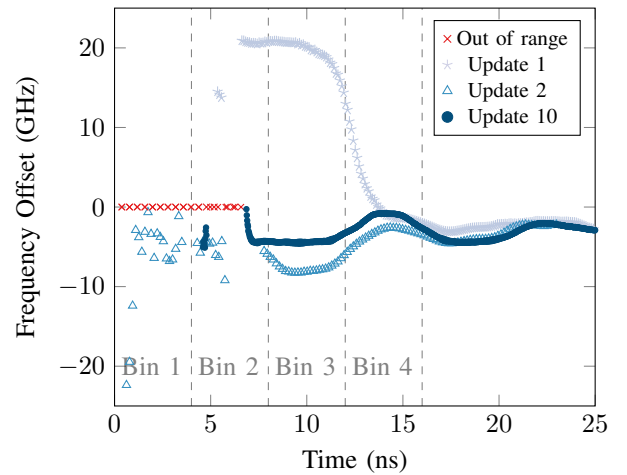}
\caption{Pre-emphasis weight calculation for a large rear section current swing of 45~mA. Each trace is averaged over 16 consecutive bursts. Each sample weight updates based on the frequency offset measured within its bin. }
\label{fig:rearfilter}
\end{figure}

\begin{figure}
  \centering
  \input{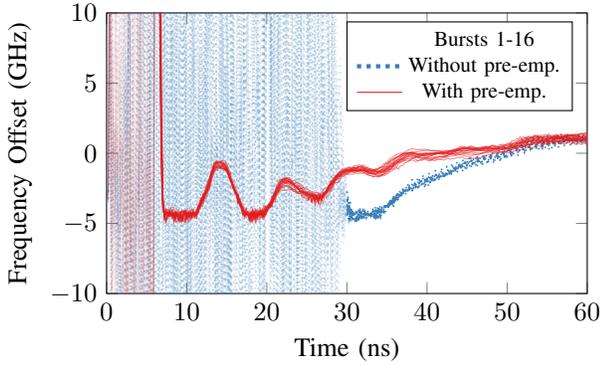}
\caption{16 consecutive bursts for a rear section current swing of 45~mA, with and without pre-emphasis. Pre-emphasis weights were calculated by the optimiser, reducing the switch time from 30.3 to 7.1~ns.}
\label{fig:before_after}
\end{figure}

Fig.~\ref{fig:rearfilter} shows the algorithm's execution for a switch with a large rear current swing of 47 to 2~mA. Each trace plots the instantaneous frequency offset of $\lambda_2$ averaged over 16 consecutive bursts. The optimiser quickly reduces the instantaneous offset to within $\pm$10~GHz, then more gradually reduces the offset over the next 9 updates.
The optimiser's final result is shown in more detail in Fig.~\ref{fig:before_after}, where 16 consecutive bursts are displayed individually, with and without pre-emphasis. The optimised pre-emphasis weights reduce the switch time from 30.3~ns to 7.1~ns, with repeatable burst-to-burst performance. The oscillations in instantaneous frequency offset are due to the low sample rate (250 MS/s) of our AWG which exhibits Fourier components in its square wave driving signals.   

\section{Phase Pre-emphasis}

\begin{figure}
  \centering
  \input{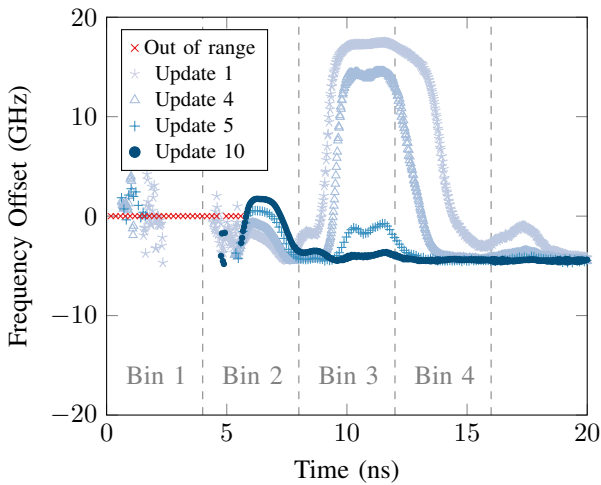}
\caption{Mode hop suppression using pre-emphasis optimisation on the laser phase section. Each curve shows the average frequency offset over 16 bursts.}
\label{fig:phasefilter}
\end{figure}

A common challenge facing fast switching semiconductor lasers is that, when stabilising on a target channel after a large current swing, the primary lasing mode can temporarily ``mode hop" to a neighbouring mode \cite{Maher2012EOCC}. Fig.~\ref{fig:phasefilter} Update 1 shows the instantaneous frequency offset of a 25~mA rear switch, after application of Algorithm~\ref{alg:filter}. We observe that although the lasing wavelength arrives at its target after 6~ns, a mode hop occurs between 9 and 14~ns. We found that applying pre-emphasis to just the rear section was insufficient to correct for mode hops; applying greater pre-emphasis on samples 3 and 4 would cause the laser to significantly overshoot the target, taking tens of nanoseconds to recover. 

Instead, the mode hop can be compensated for by applying an optimiser to the DS-DBR laser phase section. As with Algorithm~\ref{alg:filter} we used the bin-based feedback method to update each tap, but now using the simpler update term $\bm{y}=\bm{h}\bm{x}$ (substituting line 3 of Algorithm~\ref{alg:filter}). Fig.~\ref{fig:phasefilter} shows the progression of this 4 sample optimiser after 4, 5 and 10 updates. We observe that the optimiser converges nonlinearly, but is able to correct the mode hop without any manual adjustments.   

\section{Results}

\begin{figure}
  \centering
  \noindent\includegraphics[width=\linewidth]{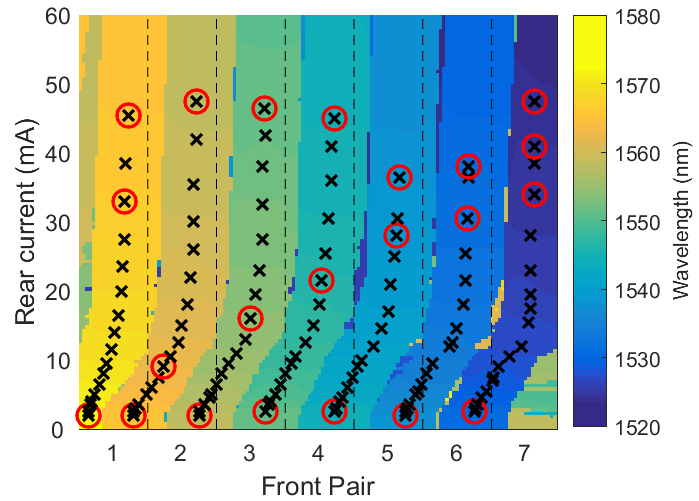}
\caption{DS-DBR laser tuning map. The positions of 122$\times$50~GHz ITU channels are marked as crosses. The 22 channels under test are circled.}
\label{fig:map}
\end{figure}

We tested the regression optimiser on a selection of the most challenging switching combinations available on the DS-DBR laser. To select these, we swept the laser rear and front grating currents and measured the resulting wavelengths on an optical spectrum analyser. This is shown in Fig.~\ref{fig:map}.
The lowest (190.65~THz, 1572.476~nm) and highest (196.70~THz, 1524.110~nm) accessible International Telecommunication Union (ITU) frequency channels were selected, covering 6.05~THz. 
This range supports 122$\times$50~GHz spaced ITU channels; these positions are marked as crosses in Fig.~\ref{fig:map}. Phase grating currents were tuned to minimise the maximum rear grating to 47.5~mA and ensure the static DS-DBR channels were within 300~MHz of their target frequencies. To test the most difficult switches, the lowest and highest rear current channels on each front section were selected, along with 8 extra channels. This forms a test set of 22 channels, circled in Fig.~\ref{fig:map}, which cover the extremes of frequency, necessary current swings and all lasing sections. 

\begin{figure}
  \centering
  \input{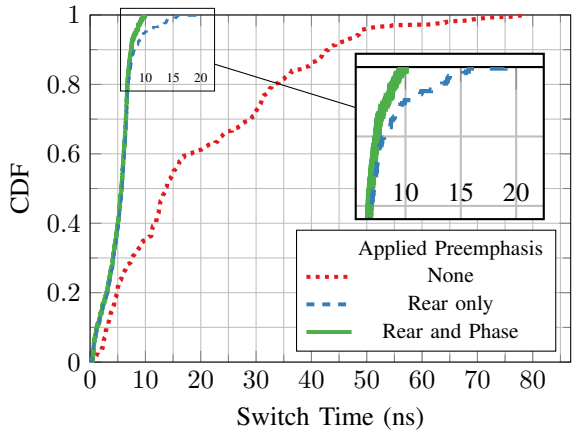}
\caption{Cumulative distribution of the time taken to reach $\pm$10~GHz of the target wavelength. 462 worst case transitions are tested. Applying pre-emphasis on the rear and phase sections bring all switch times under 10~ns.}
\label{fig:cdf}
\end{figure}

Algorithm~\ref{alg:filter} was applied to all $22\times21=462$ worst case switching combinations. Fig.~\ref{fig:cdf} shows the cumulative distribution of the time taken to reach $\pm$10~GHz of the target wavelength. By applying the regression optimised weights to just the rear section, 95\% of switches are complete within 10~ns, with all switches complete within 20~ns. The 25 cases that exceeded 10~ns almost all experienced mode hops. We then applied regression optimised pre-emphasis to the phase section for just these cases, after which they were all brought below 10~ns. The worst case switch time was left at 9.7~ns. \correction{After 15~ns, all switches were within $\pm$5 GHz of their target wavelength; this has previously been shown to be sufficient for burst mode coherent detection \cite{Simsarian2014}.}

\begin{figure}
  \centering
  \input{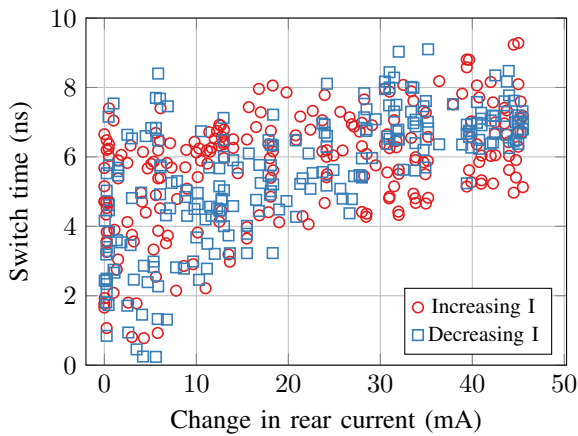}
\caption{Average switch time measured by beat intensity versus absolute change in rear section current, I. Variation in front section switching speed contributes to the 4~ns spread observed throughout the data.}
\label{fig:dIvsT}
\end{figure}

Fig.~\ref{fig:dIvsT} plots the average switch time versus the absolute change in rear section current. The $\sim4$~ns variance measured throughout this data can be attributed to the response times of the digital-select analogue-route front current switches, which varied between 3 and 7~ns. Given this, the data is well grouped and shows a positive correlation between switch time and current, indicating that the optimiser applied in this work has brought the DS-DBR laser's switch transitions close the limits set by the electron carrier mobility \cite{Simsarian2006,Simsarian2014,Funnell2017}.  

\begin{figure}
  \centering
  \input{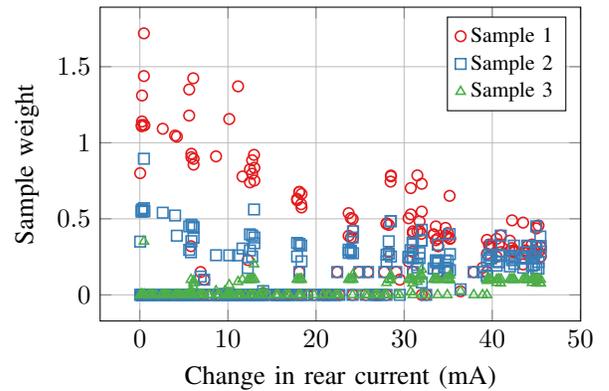}
\caption{Final rear section pre-emphasis weights as calculated by the regression optimiser. Sample weights show a progression but are not predicable. }
\label{fig:fallTaps}
\end{figure}

Fig.~\ref{fig:fallTaps} shows the final sample weights for all switches with decreasing rear currents (rising currents showed similar results). As expected, the first and second sample weights are the most important, applying significant pre-emphasis over the first 8~ns of the switch. The third weight is $\sim0.1$ or less, providing smaller stabilisation corrections. The fourth weight is set to zero in almost all cases so is excluded from this plot.
Each sample weight shows a progression but is not predictable, and in some cases no pre-emphasis is required at all. This is reflective of the complex interaction of the laser's multiple lasing sections and common grounding plane, and highlights the need for case-specific pre-emphasis calculation.    

\section{Conclusion}

A linear regression optimisation algorithm has been applied to a multi-section semiconductor laser to achieve fast wavelength switching.
The optimiser reliably calculates the pre-emphasis sample weights required to reduce the switching time in a commercially available DS-DBR laser without any manual adjustments. By testing this algorithm on a representative sample of worst case switching transitions, we demonstrate any-to-any switching within 10~ns over a switching bandwidth of 6.05~THz (122$\times$50~GHz channels). 
\correction{The fast, wideband switching demonstrated here can help improve the throughput and scale of optically switched wavelength routed networks.}
The regression algorithm is suitable for in-situ, hands-free calibration of the pre-emphasis sample weights in a fast switching transceiver system.





%


\end{document}